\documentclass[12pt]{iopart}

\usepackage{graphicx}
\usepackage[utf8]{inputenc}
\usepackage[T1]{fontenc}
\usepackage{comment} 
\begin{document}  

\title{"Cargo-mooring" as an operating principle for molecular motors}
\author{Bartosz Lisowski and Michał Żabicki}
\address{M. Smoluchowski Institute of Physics, Jagiellonian University, Reymonta 4, 30-059 Krakow, Poland}
\eads{\mailto{bartek.lisowski@uj.edu.pl}, \mailto{michal.zabicki@uj.edu.pl}}

\maketitle

\begin{abstract}

Routinely navigating through an ever--changing and unsteady environment, and utilizing chemical energy, molecular motors transport the cell's crucial components, such as neurotransmitters and organelles.
They generate force and pull cargo, as they literally walk along the polymeric tracts, e.g.\ microtubules.
However, using experimental data one may derive that the energy needed for this pulling would take the most part of the $22$ $k_BT$ that ATP hydrolysis makes available.
In such a case there would not be sufficient energy left to drive the conformational changes in the catalytic cycle of the protein.
Furthermore, the medium inside living cell is viscoelastic.
Pulling cargo in such an environment takes more energy than in aqueous buffer solution.
Here we propose a mechanism for the motor to more efficiently utilize chemical energy.
In our model the energy is used to ratchet the cargo forward.
The motor no longer pulls, but only holds a bead or a vesicle, allowing for Brownian motion in a range limited by the elasticity of the motor--cargo--track system.
The consequence of such a mechanism is the dependency of motion not only on the motor, but also on the cargo (especially it's size) and on the environment (i.e.\ it's viscosity and structure).
However, current experimental works rarely provide this type of information for in vivo studies.
We suggest that even small differences between assays can impact the outcome.
Our results agree with those obtained in wet laboratories and provide novel insight in the mechanism of a molecular motor's functioning.
\end{abstract}


\section{Introduction}
It is well understood that eukaryotic cells can not rely on free diffusion, which is simply too slow and too uncontrollable to fulfill the transportation needs. 
For this reason the motion of certain vesicles and organelles involves motor proteins and their filamentous tracks.
This guarantees proper cell functioning. 
Intracellular active transport is something that every eukaryotic cell has to coordinate, maintain and constantly shape.
It involves many players, like different types of filaments, molecular motors and cargoes that are moved from one place to another. 
Some elements can be studied in isolation, {\it in vitro}.
Nevertheless, much about motors and how they work is still unknown.

Milestone works on kinesin-1, experimental \cite{SVOBODA:1993uk,SVOBODA:1994vn,Howard:1997tx}, as well as theoretical \cite{Astumian:1994tk,Astumian:1997be,Julicher:1997zz} have brought many scientists from different fields to the topic of active intracellular transport. 
A lot of questions have been both asked and answered.
We now know for instance that the {\it walking} pattern is "hand-over-hand" and not "inchworm" \cite{Asbury:2003ff,Yildiz:2004uo}.
Also the role of the neck linker that connects the two heads has been cleared up to a large extent \cite{Kozielski:1997wo,Tomishige:2006gm,Shastry:2011wp}.
We have attained the good understanding of a relation between mechanics and chemistry \cite{Mori:2007jy}.

But there are still uncertainties.
The moving motor protein uses the chemical energy to overcome the viscous friction of the cytosol and to drive the conformational changes in the catalytic cycle.
In {\it in vitro} experiments, moreover, it is possible to apply an external load with an optical tweezer.
An external load that the motor has to also work against.

The drag force $F_\gamma$ is given by $F_\gamma = \gamma  v$, where $\gamma$ is the drag coefficient and $v$ is the average velocity of a probe.
The observed average velocity of kinesin-1 is about $800$ nm$/$s, corresponding to about one hundred $8$ nm steps per second.
The $L= 8$ nm represents the periodicity of the microtubule as well as the kinesin's step length.
Upon closer inspection it appears that the actual mechanical $8$ nm step is completed in $T = 15$ $\mu$s, i.e.\ roughly $0.1$ percent of the about $10$ ms it takes to complete the entire catalytic cycle \cite{Carter:2005bf}. 
The energy needed to overcome the friction can be evaluated as follows.
For a spherical cargo of radius $R = 250$ nm, as in \cite{Carter:2005bf}, moving in an environment having the viscosity of water $\eta_{H_2O}=10^{-3}$ Pa s one can write the Stokes' law:
\begin{equation}
F_\gamma = 6  \pi  \eta_{H_2O} R  \frac{L}{T}.
\label{Stokes}   
\end{equation}
The energy $E$ needed for a mechanical step is given by $E=F_\gamma L = \gamma v L$.
In the discussed case it is equal to:
\begin{equation}
E= 6 \hspace{0.7 mm}  \pi  \eta_{H_2O} R  \frac{L^2}{T} \approx 4.9 \mbox{ } k_BT.
\label{Energy}   
\end{equation}
Here we have used the thermal energy unit $k_BT \approx 4.1\times10^{-21}$ J in the temperature of $300$ K.
Taking the viscosity of water may be inaccurate and lead to an underestimated result.
Surface effects may be significant in the actual experiments, as discussed in \cite{Beausang:2007da} (also see \Sref{Crow}).

In optical tweezer experiments, the external load $F_L$ can be exerted on the cargo. 
As mentioned before, the kinesin-1 molecule then pulls the cargo to overcome both the viscous drag and the load force.
With an added load force $F_L$, \Eref{Energy} becomes:
\begin{equation}
E= \left(6   \pi  \eta_{H_2O}  R  \frac{L}{T}+F_L\right)L.
\label{Energy_load}   
\end{equation}
The range of $F_L$ in which the motor can still operate (that is, one can observe directed motion of a bead) is between $0 - 7$ pN.
For the load $F_L = 6$ pN  we have (after the unit conversion) $E\approx 16$ $k_BT$ for a bead of $R = 250$ nm.
For the chemical transitions in a catalytic cycle to be irreversible, they each have to be driven by an energy difference of about 2 $k_BT$.
This means that about $10$ $k_BT$ is necessary for the chemical part of the entire cycle \cite{Bier:2008te,Cross:2004cv}.

At physiological conditions the hydrolysis of one ATP molecule releases an energy of about $22$ $k_BT$.
This energy must be sufficient to provide for both the mechanical and the chemical part of the cycle.

As shown above, this is not enough to cover the $(10 + 16)$ $k_BT$ energy expense derived from the generally accepted paradigm that kinesin-1 pulls the cargo.
Bigger beads, having a radius of around $R = 500$ $\mu$m, have also been used \cite{SVOBODA:1993uk,SVOBODA:1994vn,Carter:2005bf}.
For these the energy needed to overcome viscous friction is higher.

The calculations presented above set the lower limit for the {\it in vitro} dynamics of a single kinesin motor.
Any modification of this scheme requires more energy input.
In their work, Holzwarth {\it et al.} estimated the energy needed for pulling the cargo in the buffer solution and in the cytoplasm \cite{Holzwarth:2002vc}.
Using the data from \cite{Nishiyama:2001bk} they concluded that, while active motion in an {\it in vitro} buffer solution needs less energy than is provided by ATP hydrolysis, the situation changes drastically in a viscoelastic cytoplasm. 
Basing our reasoning on more recent data from \cite{Carter:2005bf} we even more challenge the accepted model of kinesin pulling the cargo not only {\it in vivo}, but also {\it in vitro}.

Instead, we propose a model of a molecular motor working as a {\it  mooring rope} which sequentially changes the docking point while walking along its track. 
We show that a motor having the features of the kinesin-1 may ratchet the diffusion of the cargo.
Obtained results agree with experimental data and the construction of the presented model allows for a new insight into the functioning of motor proteins.

\section{Model}
\label{model_presentation}
\begin{figure}
\begin{center}
\includegraphics[width=0.9\textwidth]{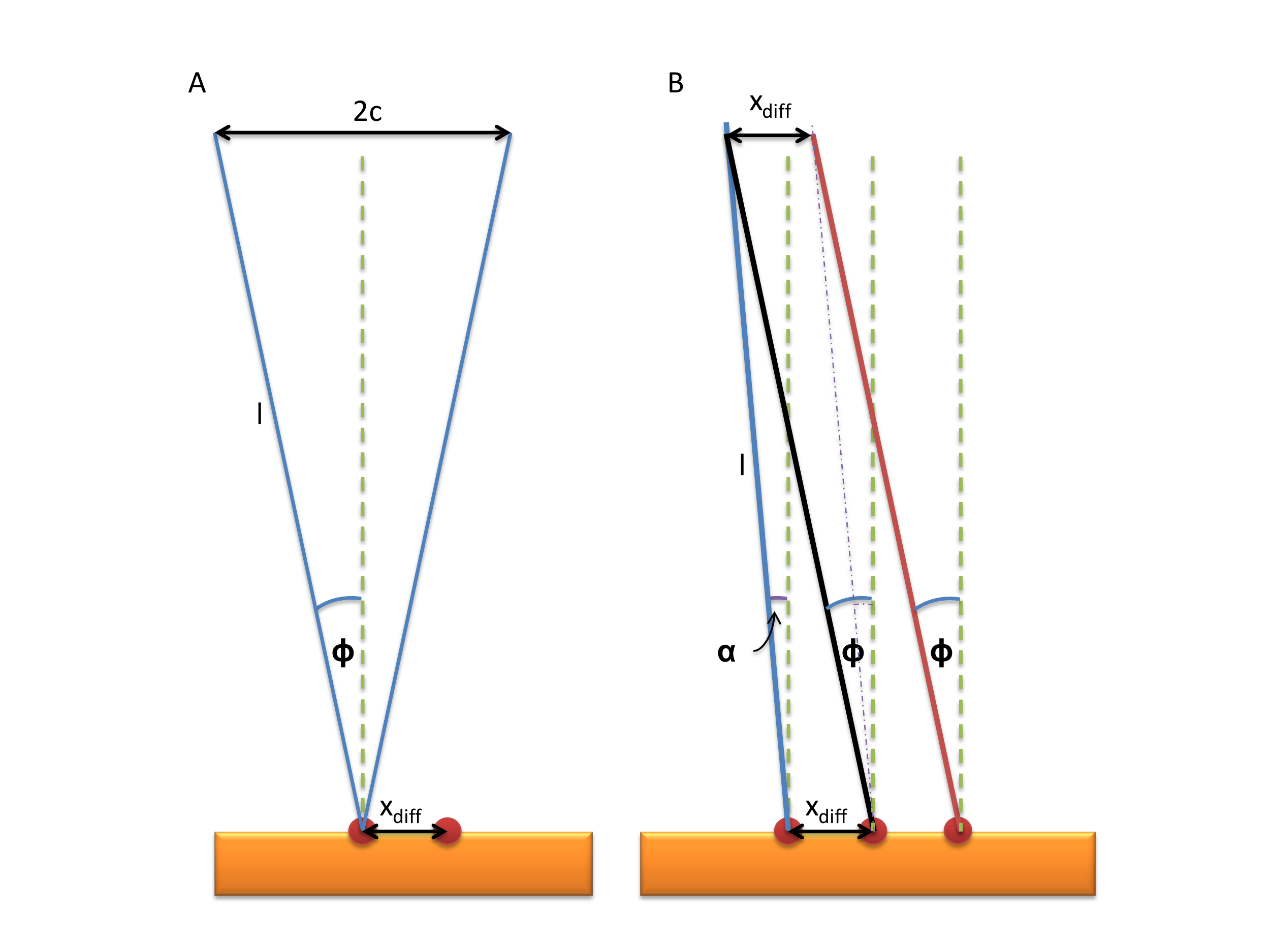}
\caption{{\bf Cartoon presenting the model's idea}. 
The cargo (not shown here) is linked to the track by a rod of length $l$.
The red dots are the binding sites.
The rod may wiggle left or right due to Brownian motion of the cargo.
The angle $\phi$ represents the maximum deflection from the vertical orientation. ({\bf A}). 
We assume that the rod makes a step of length $r$ when the angle of deflection equals $\alpha$.
At a new binding site the initial angle of deflection equals $\phi$.
From there it has to wait until the cargo will again diffuse by a distance $r$, so that the angle of deflection will once again be equal to $\alpha$. ({\bf B}).}
\label{cartoon_model}
\end{center}
\end{figure}

We start with the motor as a uniform rod of length {\it l}, see \Fref{cartoon_model}. 
It links the cargo --- a spherical bead with radius $R$ --- with the track.
All of those elements are immersed in a buffer solution of known viscosity. 
The track has a periodic structure, with special domains --- binding sites --- positioned every $8$ nm. 
This corresponds to the known molecular structure of microtubule \cite{Howard:2001wk}. 
Generally, the motor moves from one binding site to the next.
The stepping process is coupled to a sequence of a chemical reactions. 
For these reactions the motor takes substrates from the buffer solution and acts like an enzyme. 
In our model conformational changes let the rod detach from the previously occupied site and reattach at the neighboring one. 
The motion is conceived as taking place in one dimension and we allow the motor to move only from the left to the right.

The cargo is subject to free diffusion in the surrounding solution. 
It is attached to the motor, which in turn holds on to the binding site, hence it may move only within a limited range.
This range is determined by the motor's length and the elasticity of the motor--track connection.
Due to thermal motion, between steps the motor swings back and forth around the docking point.
The angle, however, will not be bigger than a maximum deflection,~$\phi$.
This is mindful of a windswept balloon on a string that is tight to a pole.

As depicted in \Fref{cartoon_model} {\bf B}, the motor may change its docking point, i.e.\ take an $8$ nm jump from the left to the right (corresponding to kinesin's forward step, which means a step taken into the direction of microtubule's {\it plus} end) when deflected by an angle $\alpha$ from the vertical. 
After transition, the motor is attached to the subsequent binding place on the track, deflected by the maximal angle $\phi$. 
Here, to make another step, the motor has to wait for the cargo to diffuse again.
The distance $x_{diff}$, that needs to be covered by the cargo's diffusive motion to enable the motor to perform another step, is equal to the distance between two neighboring binding sites, i.e.\ $8$ nm in the case of kinesin-1 walking along microtubule. 
We denote this distance by $r$.
When doubled, it will give the full range available for the cargo's diffusive motion around each binding site (see \Fref{cartoon_model}).
Distinction between $r$ and  $x_{diff}$ is important for the results presented in \Sref{compar}.
When those quantities are equal, we use the notation $r$.

In the simulation the position of the bead $x$ evolves in time as \cite{Beausang:2007da}:

\begin{equation}
x(t+\Delta t)=x(t)+\sqrt{2D\Delta t}\mbox{ }\xi(t)-\frac{F_L(x(t))}{\gamma}\Delta t,
\label{LangEq}   
\end{equation}
where $\xi(t)$ is uncorrelated Gaussian random noise with zero mean and a standard deviation of one.
$D$ is a diffusion constant, related to the drag coefficient $\gamma$ through the Stokes-Einstein relation:\ $D=\frac{k_B T}{\gamma}$.
$F_L(x(t))$ is a load force exerted by an optical tweezer. 
The latter we may write in an explicit form as a hookean-spring equation: 
\begin{equation}
F_L(x(t))=\kappa x(t),
\label{force}   
\end{equation}
where $\kappa$ is the tweezer's stiffness. 
\Eref{LangEq} can now be rewritten in slightly different form: 
\begin{equation}
x(t+\Delta t)=x(t)+\sqrt{2D\Delta t}\mbox{ }\xi(t)-\kappa x(t)\frac{D}{k_BT}\Delta t.
\label{eqfull}   
\end{equation}

Since for a spherical bead of radius $R$ moving in a liquid of viscosity $\eta$ we can write $\gamma = 6\pi \eta R$ and assume the applicability of the Stokes-Einstein relation, we can calculate the diffusion coefficient of the cargo. 
Taking the radius of the bead $R$ = $0.28$ $\mu$m, as in \cite{Carter:2005bf},  and using, for reasons discussed in \cite{Beausang:2007da}, an {\it effective} viscosity of the buffer solution $\eta$ = $2.4$ $\times$ 10$^{-3}$ Pa~s, we obtain $D \approx 1.6 \times 10^5$ $nm/ s^2$ at a temperature of $300$ K. 
From \cite{Carter:2005bf} we take the trap stiffness coefficient $\kappa$ = $0.065$ pN~nm$^{-1}$.

Since, for kinesin-1, the conformational change involving the mechanical step lasts around $15$ $\mu$s (for a bead of radius $R=250$ nm \cite{Carter:2005bf}), we take as the time step in the simulation half of this value $\Delta t$ = $7.5$ $\mu$s.
Such a time step should be sufficiently small to catch all the dynamical phenomena discussed here.

\section{Results}

\subsection{The molecular motor can not walk faster than it can}

The initial results of our model show that allowing the molecular motor to make a step whenever the cargo position allows is not enough to obtain results that agree with experimental data. 
The resulting motion is unrealistically fast for small load forces. 
This occurs because of the very rapid diffusion; the second term in \Eref{eqfull} dominates both other terms. 
Only for higher loads are results similar to those observed experimentally. 
The obvious oversight was not taking into account the time needed for hydrolysis of one ATP molecule.  
The chemical cycle is evidently not an instantaneous process. 
To deal with this problem we introduce the so-called kinesin-cycle-limiter and determine its value constant and equal to $10$ ms, which corresponds well with the waiting times for unloaded kinesin as measured in \cite{Carter:2005bf} (see Figure\ 2b therein).
In our simulations no step is possible in time shorter than $10$ ms.
This value is then the minimal dwell time of the motor. 

It is interesting to note that, while the chemomechanical approach towards modeling motor proteins \cite{Liepelt:2007ko,Clancy:2011if} considers the impact of external forces on reaction rates through Arrhenius' law, our approach gives a simple physical mechanism behind the force-velocity relation.
Even for a fixed limiter, unaffected by external forces, we observe the decrease in the motor's velocity in the presence of high loads. 
In this regime the duration of the chemical cycle is no longer the limiting factor --- it is the diffusion that is affected. 

\subsection{Velocity}
For each of the conditions described in the following sections, we calculate the dwell time in which the diffusing cargo covers at least $8$ nm in the direction of the motor's motion. 
This would enable a docking point change.
We calculate the motor's average velocity by dividing the $8$ nm by the dwell time.
The actual $15$ $\mu$s in which the movement takes place is negligible compared to this dwell time.
We also consider diffusion distances of $x_{diff}=4$ nm and $x_{diff}=16$ nm, as depicted in \Fref{longC}~A.
There we divide the range $x_{diff}$ of the size $4$ nm, $8$ nm (as is the standard way) and $16$ nm by the corresponding waiting times between $8$ nm-long steps of the motor. 
The question we then ask is: if the motor could make a step, not after the cargo diffuses $8$ nm, but after $4$ or $16$ nm, what would change?
It is a legitimate question, as we do not know the range size $x_{diff}$.
Ultimately we can obtain an estimate of the length $x_{diff}$ by comparing theoretical force--velocity curves with experimentally observed ones.

\subsection{Trajectory}

\begin{figure}
\begin{center}
\includegraphics[width=0.9\textwidth]{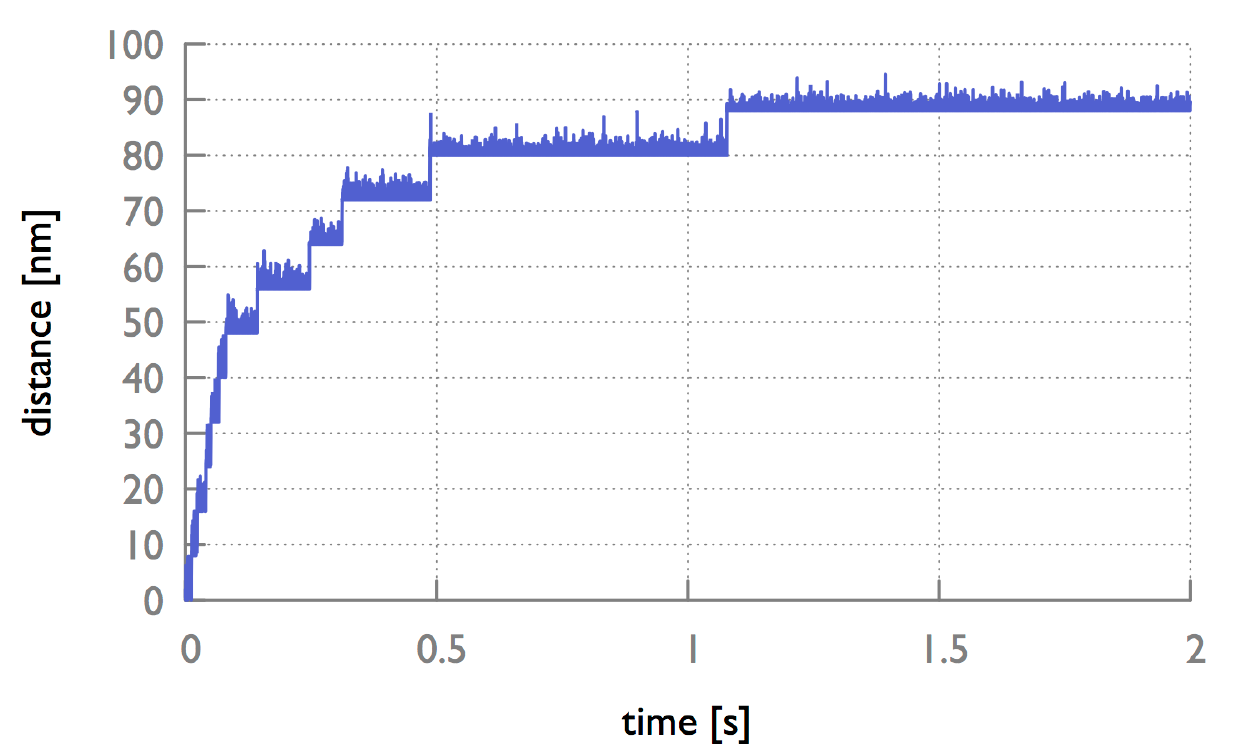}
\caption{{\bf Example of a model trajectory.} 
Cargo has been attached to the simulated optical-trap spring at $x=0$. 
Through the motor it travels $80$ nm away from the trap center in less than half a second. 
The full $88$ nm corresponds (via the trap stiffness $\kappa$) with load force $F_L$~$\approx$~$6$ pN. 
This plot compares favourably to the experimental results from \cite{Carter:2005bf}. 
Model parameters are: $\eta=2.4\times10^{-3}$ Pa s, as in \cite{Beausang:2007da} and $R=560$ nm and $\kappa = 0.065$ pN nm$^{-1}$, as in \cite{Carter:2005bf}. 
The time step is $\Delta t=7.5\times10^{-6}$ s and the range is $r=8$ nm.}
\label{trajectory}
\end{center}
\end{figure}

To simulate the optical trap behavior, the motor with attached cargo has been placed at position $x=0$. 
The same position has been assigned to the center of the simulated optical trap. 
Over the course of time, cargo --- subject to diffusive motion --- reaches the critical point (here: $8$ nm).
At the critical point the motor changes position and allows cargo to diffuse in a new range.
The further it moves away from the point $x=0$, the greater is the load force exerted by the optical trap.  
However, with the help of the motor, acting like a mooring rope which holds the ship near a bollard, cargo can cover large distances.
\Fref{trajectory} depicts the results of the simulation of these procedure. 
It compares well with experimental results obtained by Carter and Cross (Figure~1 in \cite{Carter:2005bf}). 
The cut-offs on the bottom of each range are because of the reflecting barrier.

\subsection{Cargo dimension}
\label{carg}

\begin{figure}
\begin{center}
\includegraphics[width=0.9\textwidth]{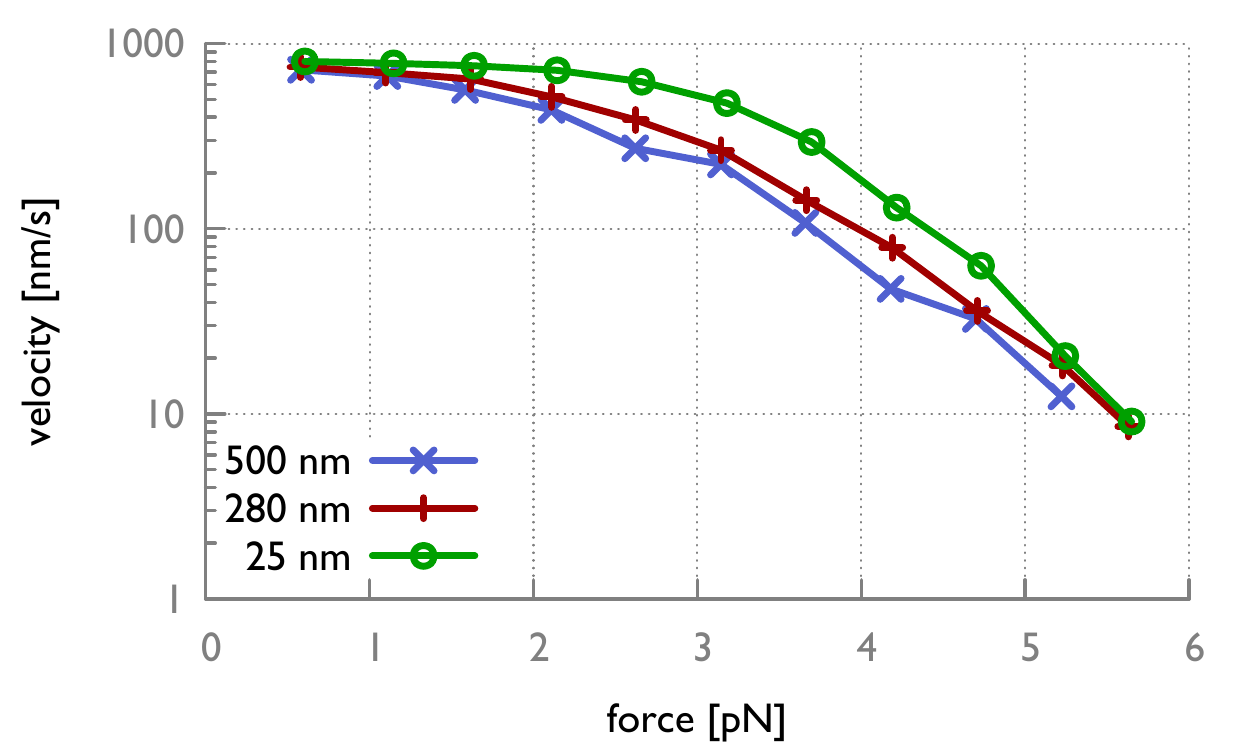}
\caption{{\bf Velocity as a function of load force $F_L$ for different values of the cargo radius.} 
Model parameters are: $\eta=2.4\times10^{-3}$ Pa s, as in \cite{Beausang:2007da} and $\kappa=0.065$ pN nm$^{-1}$, as in \cite{Carter:2005bf}. 
The time step is $\Delta t=7.5\times10^{-6}$ s and the range is $r=8$ nm.}
\label{longR}
\end{center}
\end{figure}

As our base cargo's radius size we have used $R = 280$ nm, which corresponds with the size of the beads used by Carter and Cross in \cite{Carter:2005bf}.
Additionally, we have also analyzed motor--cargo dynamics for smaller ($R = 25$ nm) and bigger ($R = 500$ nm) cargoes. 
The simulation results are presented in \Fref{longR}.
With $\gamma=6\pi \eta R$ it is obvious that the drag is larger for larger beads. 
The diffusion coefficient $D=\frac{k_BT}{\gamma}$ should be smaller for larger beads.
For more crowded environments the relations between drag, diffusion and bead size may be more complicated \cite{OchabMarcinek:2011gt,Kalwarczyk:2011fa}.

\subsection{Viscosity}
\label{Crow}

\begin{figure}
\begin{center}
\includegraphics[width=0.9\textwidth]{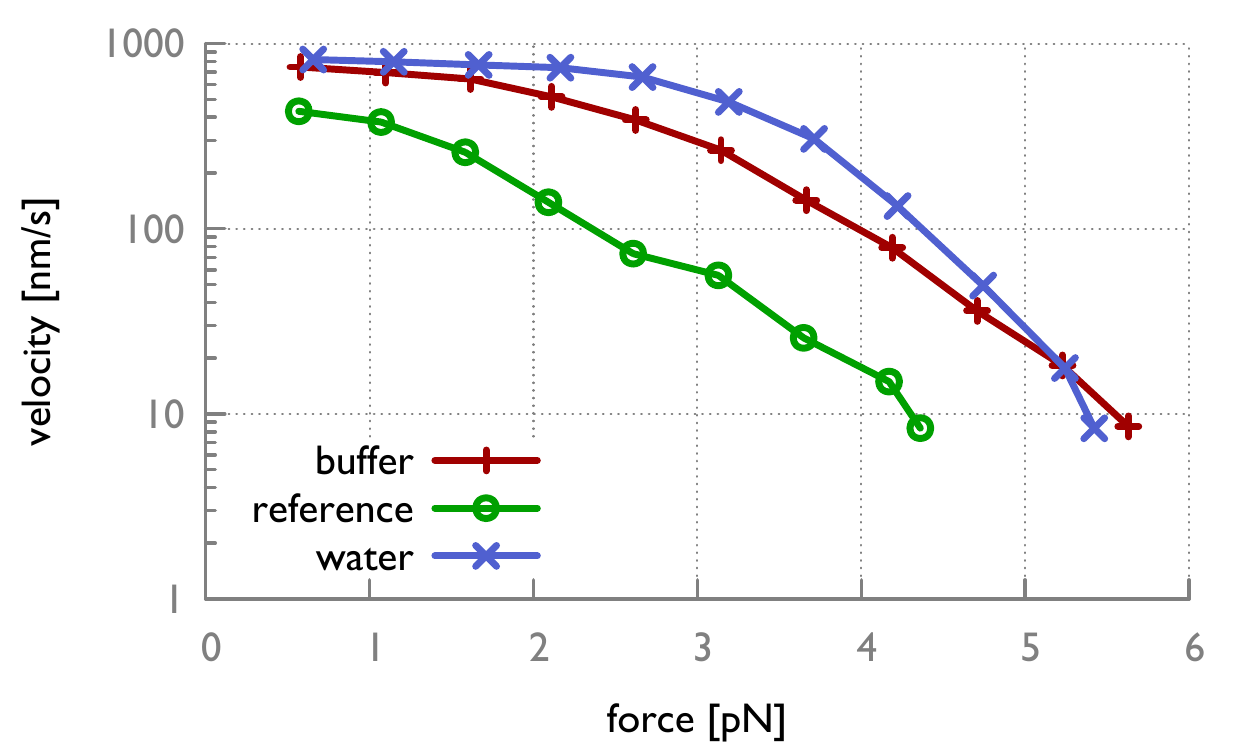}
\caption{{\bf Velocity as a function of load force $F_L$  for different viscosities.} 
Model parameters are: $R=560$ nm and $\kappa=0.065$~pN~nm$^{-1}$, as in \cite{Carter:2005bf}. 
The time step is $\Delta t=7.5\times10^{-6}$ s and the range is $r=8$ nm. 
Simulations for different solutions (buffer with effective viscosity  $\eta = 2.4 \times10^{-3}$~Pa~s, as in \cite{Beausang:2007da}, water with it's normal viscosity  $\eta_{H_2O} \approx 1 \times10^{-3}$~Pa~s, and some reference solution with viscosity  $\eta_{\mbox{ref}}=24 \times10^{-3}$~Pa~s) reveal that higher viscosity generally leads to lower speed.
For details see text.}
\label{longD}
\end{center}
\end{figure}

Most of the molecular motor experiments are conducted {\it in vitro}, i.e.\ in some buffer solution of a homogenous viscosity.
In this study we examine motion in three homogeneous solutions --- buffer solution, water and a reference solution  (\Fref{longD}). 
For the buffer we took the effective viscosity  $\eta=2.4\times10^{-3}$~ Pa~ s, as in \cite{Beausang:2007da}. 
The presence of the wall, resulting from the experimental procedure, as well as the interactions with motor makes that the cargo does not "feel" the real viscosity of the buffer, which is of the order of the viscosity of water, $\eta_{H_2O}\approx1\times10^{-3}$~ Pa~ s, but some higher viscosity.
The third solution we used is a hypothetical environment with a viscosity ten times higher than the $\eta=2.4\times10^{-3}$~ Pa~ s. 
It should be pointed out that this higher viscosity can not be identified with a cell's interior.
A cell's interior is generally not homogenous and it is not merely viscous, but rather viscoelastic.
Results depicted in \Fref{longD} show the sensitivity of the velocity to the viscosity. 
As a general rule, with the same motor, the cargo's motion slows down as viscosity increases.
Higher $\gamma$ implies smaller $D$ i.e.\ smaller Brownian kicks felt by the cargo. 
As a consequence, the dwell times in one position are longer --- it takes a longer time to reach another no-return point.
Also the stall force decreases significantly for higher viscosities. 
The decrease of the amplitude of the thermal noise makes the cargo less mobile.
In spite of the damping the elastic term in \Eref{force} will then eventually dominate.
Changing the viscosity significantly affects the dynamics of the motor--cargo system.
A more sophisticated analysis probably required to describe the behavior of the motor in an inhomogenous cell environment.
Molecular crowding and elastic effects should then be included. 
Furthermore, the Stokes--Einstein relation, $D=\frac{k_BT}{\gamma}$ no longer straightforwardly applies in a viscoelastic environment \cite{Banchio:1999ed}. 
Anomalous diffusion appears common inside living cells \cite{Weiss:2003cl}.
For such diffusion, more sophisticated models are now being proposed (see \cite{Barkai:2012gg} for a short review).

\subsection{Range size and collective behavior}
\label{compar}

\begin{figure}
\begin{center}
A)\includegraphics[width=0.9\textwidth]{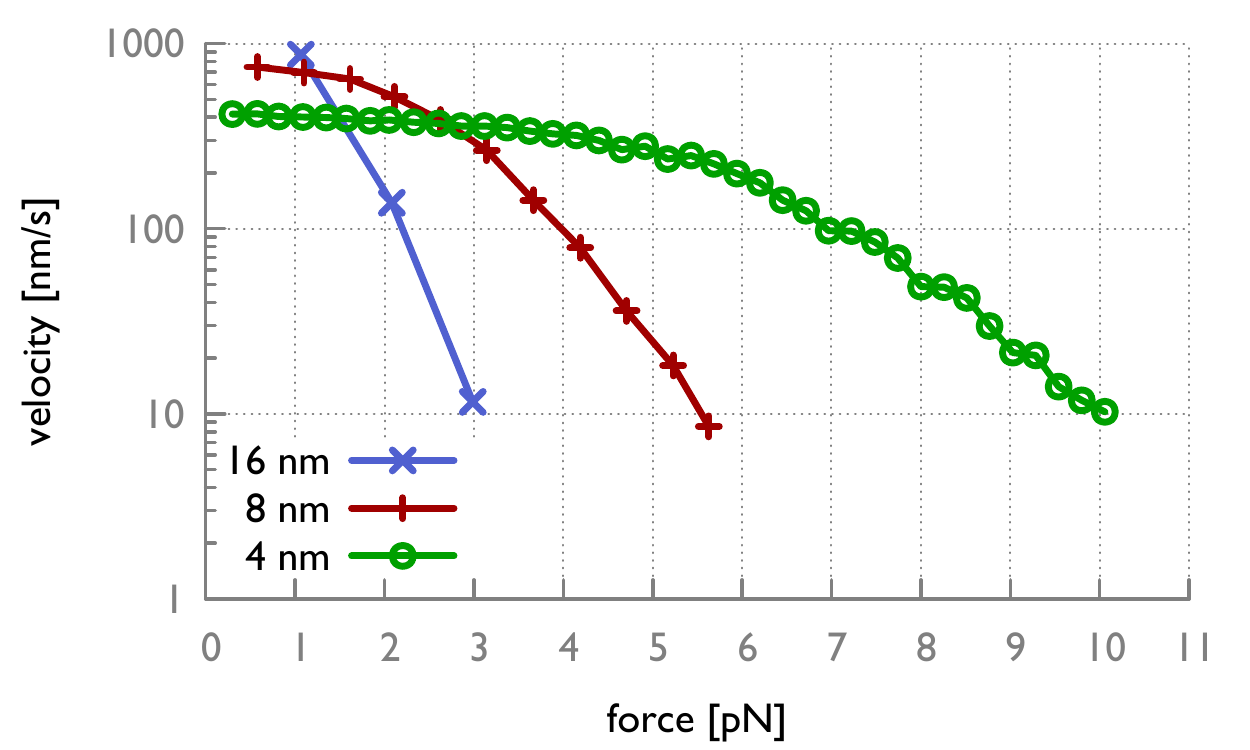}
B)\includegraphics[width=0.9\textwidth]{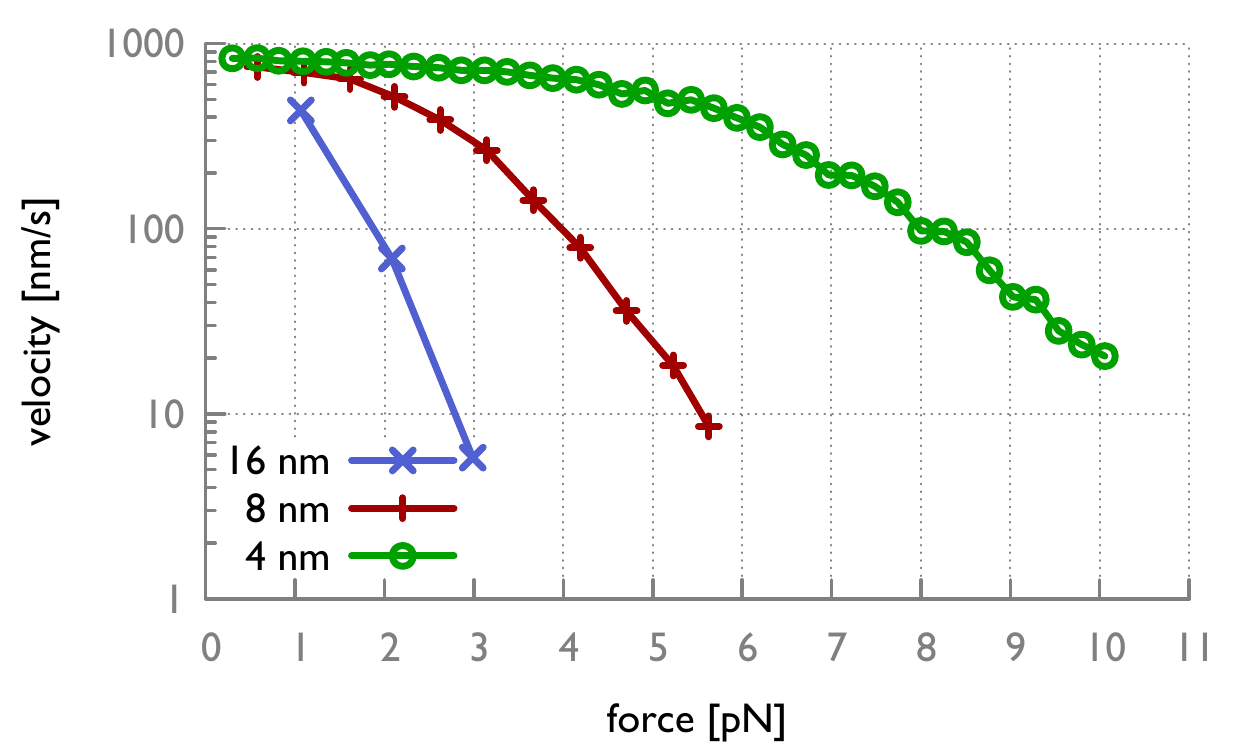}
\caption{{\bf Velocity as a function of load force $F_L$ for different ranges.} 
For smaller ranges --- i.e.\ for coupled motors working collectively --- higher values of the stall force are observed. 
Model parameters are: $\eta=2.4\times10^{-3}$ Pa s, as in \cite{Beausang:2007da} and $\kappa=0.065$ pN nm$^{-1}$, as in \cite{Carter:2005bf}.
The time step is $\Delta t=7.5\times10^{-6} $ s. Figure A --- with $v=r/\mbox{dwell time}$ Figure B --- with $v=8\mbox{ nm} / {\mbox{dwell times}}$.}
\label{longC}
\end{center}
\end{figure}

So far we have considered a range size $r$ equal to the kinesin-1 step size, i.e.\ $8$ nm. 
Below we check the behavior of the model for different values of $r$. 
The results of our simulations, as depicted in \Fref{longC}, are somehow puzzling. 
What one would expect is that a smaller range will result in faster average motion, as the cargo will achieve the no-return point faster. 
However, for $r = 4$ nm the velocity at low loads is similar to the velocity for larger ranges.
What we do see is a significantly higher stall force for the $r=4$ nm curve.

\begin{figure}
\begin{center}
\includegraphics[width=0.5\textwidth]{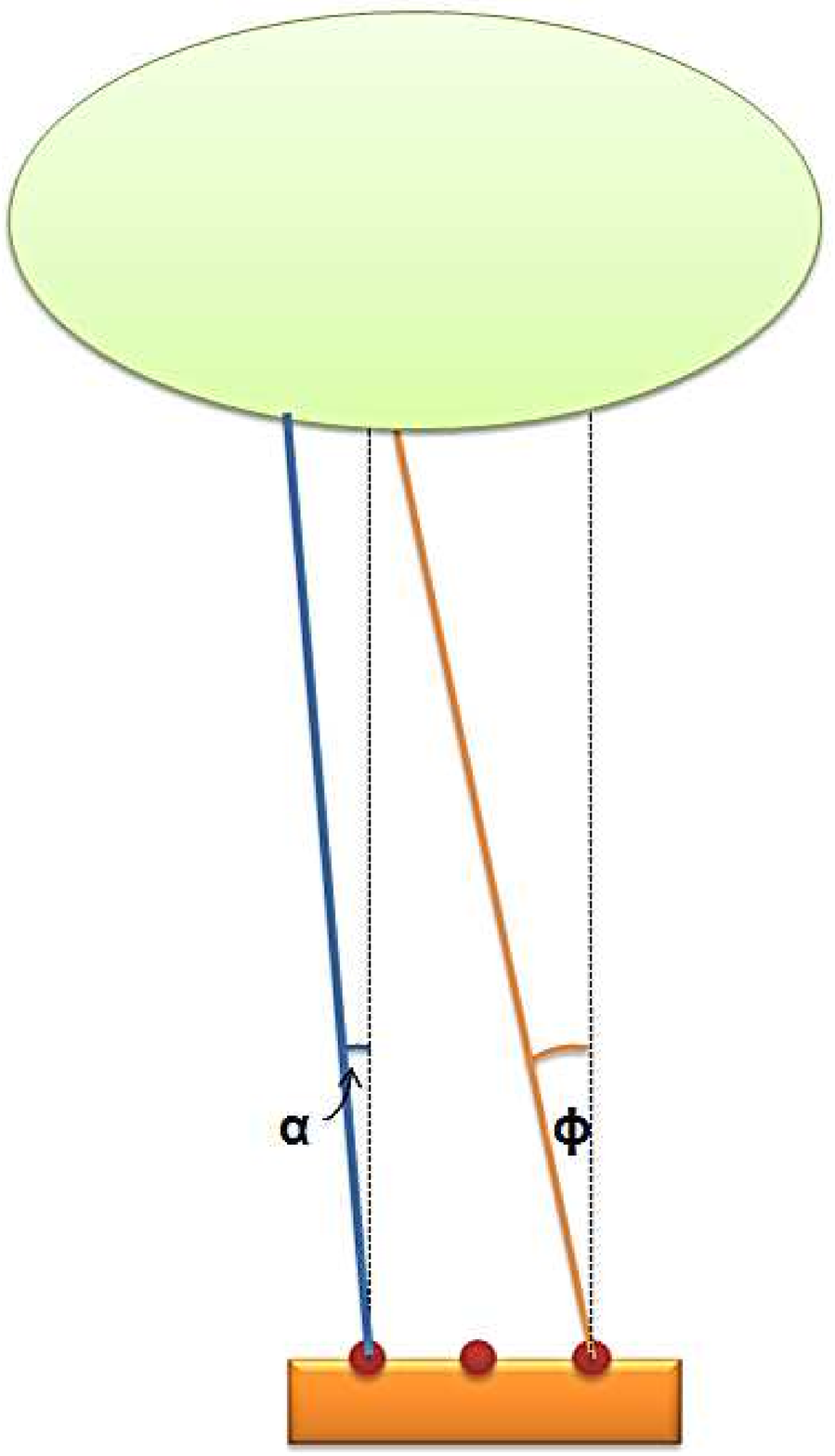}
\caption{{\bf Collectively acting motors limit the range size $r$.} 
The orange rod is maximally deflected, so that the cargo's diffusion to the left is impossible.
It can move only to the right.
All of these results in a decrease in the range over which the cargo can diffuse.}
\label{two_kin}
\end{center}
\end{figure}

It has been observed that increasing the number of motors does not lead to higher speed \cite{Shubeita:2008ek,Vershinin:2007dl}. 
Instead the system gains an ability to overcome larger loads. 
In our model, decreasing the range size $r$ may be identified with increasing the number of motors associated with one cargo. 
In this case, assuming the random location of motors, it may happen that when one motor reaches its maximum deflection $\phi$, the other can be deflected by an angle $\alpha < \phi$, as depicted in \Fref{two_kin}. 
Because of the strains arising in one molecule, the other can not increase this angle further. 
From the cargo's perspective, this is manifested in a decrease of the compartment size. 

\subsection{Dimensions}

The motion of the system has been assumed to be one dimensional.
We did not find any results indicated that kinesin-1, as used by Carter and Cross \cite{Carter:2005bf}, is side-elastic, i.e.\ that the stalk, linking the heads with the cargo, can wiggle sideways.
On the other hand, the back-and-forth elasticity is a commonly accepted phenomenon.
However, the values of the maximum deflection angles are not precisely determined \cite{Jeney:2004hc}.
Erickson et al. discuss the rotational diffusion \cite{Erickson:2011we}, that we neglect.
They show that it plays a role in the kinesin--microtubule binding dynamics in the presence of cargo.
It appears likely that for {\it in vitro} assays rotational diffusion of cargo is present, but overwhelmed by a net effect of translational diffusion and motor directionality. 
It must be noticed that for the overcrowded environment, the ratio of the rotational and translational diffusion may be different.
This is because rotational diffusion needs almost no free space around the cargo, while translational diffusion does.

\subsection{Backstepping} 

In our model we did not include the possibility of making back-steps i.e.\ jumps to the binding site on the left. 
For kinesin-1, back-steps refer to steps toward the microtubule's minus end. 
The probability of such an event is relatively low.
Carter and Cross found about one backstep for every $1000$ forward steps \cite{Carter:2005bf}.
Only for large loads do back-steps constitute a significant fraction of the total number of steps.

\section{Discussion}

Kinesin molecular motors, taking care of intracellular transport needs of eukaryotic cells, are among the best known proteins.
The energy-consuming active transport in which they are involved is thought to be an evolutionary improvement over free diffusion, the latter being just not sufficiently efficient, fast and reliable for cells bigger than $1$ $\mu$m.
However, since our knowledge about kinesins comes mainly from {\it in vitro} experiments, we are still far from understanding all the factors that impact their behavior {\it in vivo}.
Kinesin hydrolyzes one ATP molecule per step.
But how the energy is transferred into effective work is still the subject of much debate, experimentation and modeling.
Since the results of our simulations correspond well with experimental data, we suggest that, when studying the intracellular active transport, it is necessary to concentrate not only on the motor, but on the entire motor-cargo-environment system.
This is due to the fact that the viscosity of the environment significantly affects the motor's dynamics. 

{\it In vivo} kinesin pulls vesicles of about $50$ nm in diameter.
The stalk that connects the motor and the vesicle has a length of about $100$ nm.
Vesicle and/or stalk can get entangled in the cytoskeleton network. 
Kinesin that "gets stuck" in such a way will more easily detach from the biopolymer that it is walking on.
It is well known that under high loads detachment rate of a kinesin motor increases \cite{Jamison20102967}.
This means that there is a form of "communication" between the cargo and the motor --- communication that occurs via the stalk.

If the motor "feels" the impact of diffusive motion on a cargo, the generated strain may slow down or speed up chemical reactions that drive the motility.
For example, the catalytic cycle could be accelerated if the cargo can more easily find a free space to diffuse a little bit closer towards the microtubule's plus and thus trigger a next step.
This idea gives a possible explanation of the Arrhenius-like dependency of kinesin's dynamics under load.

From that the questions about kinesin waiting pattern arises: is it possible for the protein to wait for the cargo, while it reaches the position allowing for the step? 
And what conformation would it adopt? 
Answering these questions is far beyond the ability of our model. 

Mori {\it et al.} in \cite{Mori:2007jy} report on kinesin's ability to adapt different conformations facing different conditions (e.g.\ ATP concentration).
What we propose is an ability of kinesin to wait for the cargo to diffuse sufficiently far before it takes an actual step.
Our suggestion is a more extensive role for the stalk.
Strain on the stalk will not merely be a signal for detachment.
It can also trigger a next step.

\ack

We would like to thank Anna Ochab-Marcinek and Martin Bier for beneficial discussions. 
The project has been supported by the European Science Foundation within the program EPSD (Exploring Physics of Small Devices). 
BL wishes to thank the Foundation for Polish Science (International Ph.D. Projects Program co-financed by the European Regional Development Fund covering, under the agreement No. MPD/2009/6; the Jagiellonian University International Ph.D. Studies in Physics of Complex Systems).

\section*{References}

\bibliographystyle{iopart-num}
\bibliography{NEWDocak}

\end{document}